\documentclass[prl,twocolumn,showpacs,preprintnumbers]{revtex4}
\usepackage{graphicx}
\usepackage{dcolumn}
\usepackage{bm}

\def\be{\begin{equation}}
\def\ee{\end{equation}}
\def\bea{\begin{eqnarray}}
\def\eea{\end{eqnarray}}
\def\bse{\begin{subequations}}
\def\ese{\end{subequations}}

\begin{document}

\title{Quantum Computation Using Vortices and Majorana
Zero Modes of a $p_x$ + $ip_y$
Superfluid of Fermionic Cold Atoms \\
\vskip 1mm }
\author{Sumanta Tewari$^{1}$, S. Das Sarma$^{1}$, Chetan Nayak$^{2,3}$,
Chuanwei Zhang$^{1}$, P. Zoller$^4$} \affiliation{$^{1}$Condensed
Matter Theory Center, Department of Physics, University of
Maryland, College Park, MD 20742\\
$^{2}$Microsoft Project Q, Kohn Hall, University of California,
Santa Barbara, CA 93108 \\
$^{3}$Department of Physics and Astronomy, University of
California, Los
Angeles, CA 90095-1547\\
$^4$Institute for Theoretical Physics, University of Innsbruck, and\\
Institute for Quantum Optics and Quantum Information of the
Austrian Academy of Sciences, A-6020 Innsbruck, Austria }

\begin{abstract}
We propose to use the recently predicted two-dimensional
`weak-pairing' $p_x + ip_y$ superfluid state of fermionic cold
atoms as a platform for topological quantum computation. In the
core of a vortex, this state supports a zero-energy Majorana mode,
which moves to finite energy in the corresponding topologically
trivial `strong-pairing' state. By braiding vortices in the
`weak-pairing' state, unitary quantum gates can be applied to the
Hilbert space of Majorana zero-modes. For read-out of the
topological qubits, we propose realistic schemes suitable for
atomic superfluids.
%
\end{abstract}

 \maketitle


\emph{Introduction.} Topological quantum computation requires
particles that have non-Abelian statistics under interchange and
braiding. Under pairwise interchange of particle coordinates, the
many-body wave-function of particles following non-Abelian
statistics transforms via a unitary transformation in the Hilbert
space of a degenerate set of wave-functions. Such particles can
arise as the low-energy excitations of a topological phase of
matter. One
such system, which has been recently discussed in this context \cite%
{DasSarma05,Stern05,Bonderson05,Day,Bonesteel05}, is the
experimentally observed \cite{Willett,Xia} $\nu =\frac{5}{2}$
fractional quantum Hall state of a two-dimensional (2D) electron
gas, where $\nu$ is the filling fraction of the electrons. Another
promising system is a spin-triplet ($S=1$), 2D, $p_x+ip_y$
superconductor, in which certain vortex excitations have zero
energy Majorana modes\cite{Kopnin} in their cores, which endow
these vortices with non-Abelian statistics \cite{Read00,Ivanov,
Stern04}. For a spin-polarized (spinless) $p_x+ip_y$
superconductor, ordinary vortices with vorticity $N=1$ have such
Majorana modes bound at the core \cite{Read00}. For a $p_x+ip_y$
superconductor with $S_z=0$, analogous to the A phase of He 3
\cite{Vollhardt}, only the higher-energy vortices with
$N=\frac{1}{2}$ -- not the lowest energy $N=1$ vortices -- have
the Majorana modes. Nevertheless, it is possible to quench the
spin-orbit energy, which acts as a confining potential between two
$N=\frac{1}{2}$ vortices, by applying a magnetic field \cite
{Salomaa-Volovik}, and thus to make the Majorana modes potentially
realizable in experiments. Based on this appealing idea, and the
prospect that strontium ruthenate may be a quasi-2D, $S_z=0$,
$p_x+ip_y$ superconductor
\cite{Mackenzie-Review,Nelson-Science,Rice-Science}, it has been
recently proposed \cite{Sumanta} to use thin films of this
material as a system which realize non-Abelian statistics of
quasiparticles, associated with these unusual half-quantum (i.e.,
$N=\frac{1}{2}$) vortices.

Although there is nothing in principle to invalidate such a
strategy, practical difficulties may arise due to the lack of
quantum coherent motion of vortices in the films.
Moreover, since one needs to apply a threshold magnetic field
to quench the spin-orbit energy, there will be a relatively high
concentration of the half-quantum vortices, thereby rendering
independent braiding experimentally challenging. Finally, and most
importantly, since the quasiparticles of a superconductor are
chargeless, and
the Majorana modes are also spinless, there is no simple way to
couple to the state of a qubit after a braiding operation has been
performed. This makes reading out the state of the qubit
difficult. 
Hence, even though the very realization of non-Abelian statistics
through the observation of these vortices is an exciting goal in
itself, and for the purposes of topological quantum computation
several ideas to overcome the difficulties mentioned above were
recently proposed \cite{Sumanta}, it will really pay to have a
$p_x+ip_y$ superfluid system where vortex motion is likely to be
coherent, $N=1$ vortices themselves have non-Abelian statistics so
that their concentration can be independently kept low, and a
natural read-out scheme exists.

With the recent observation of a $p$-wave Feshbach resonance in
spin-polarized $^{40}$K and $^6$Li atoms in optical traps
\cite{Regal,Ticknor, Schunck}, just such a system -- an
`artificially' created $p_x + ip_y$ superfluid of spinless
fermions -- may now be within experimental reach. Exotic
non-Abelian statistics is thus tantalizingly close to fruition in
these systems. Since the atoms are in identical spin states,
$s$-wave scattering is Pauli-prohibited and a $p$-wave resonance
dominates, allowing the tunability of the atom-atom interaction in
$L=1$ channel. Recently, it has been theoretically shown
\cite{Gurarie,Cheng} that such interactions have the potential to
realize various $p$-wave superfluid states, among them,
a $p_x+ip_y$ state in the
so-called `weak-pairing' phase (chemical potential $%
\mu > 0$) in both three and two dimensions. As a function of the
Feshbach resonance detuning, which controls $\mu$, this phase
undergoes a topological quantum transition to the strong-pairing
phase ($\mu < 0$) \cite{Gurarie}. In 2D, the phase with $\mu > 0$
is topologically non-trivial because it supports zero-energy
Majorana modes at vortex cores \cite{Read00}, while, as we show
below by explicitly constructing the zero-mode wave-function, they
disappear in the topologically trivial strong-pairing phase. The
weak-pairing phase, then, is suitable for use in the hardware of a
quantum computer. Since ordinary vortices themselves exhibit
non-Abelian statistics in this case, they can be created at a low
density, allowing, in principle, independent braiding. These
vortices are also expected to be light due to the high degree of
coherence possible in optical traps, and so it is much easier to
maintain quantum coherence during the braiding operations.
Finally, as we discuss later, since the atoms, unlike the
electrons in a superconductor, have internal energy levels, this
internal structure can be manipulated to read out the states of
the qubits after the braiding operations perform the quantum
computation.

\emph{The weak and strong pairing phases and the fate of the zero
mode.} The BCS Hamiltonian for a system of spin-polarized
(spinless) fermions in a two-dimensional spin-triplet $p$-wave
superfluid state is given by,
\begin{equation}
H=\int d^2xd^2x^{\prime} \psi^{\dagger}(\vec{x}){\mathcal{H}}(\vec{x},\vec{x}%
^{\prime})\psi(\vec{x}^{\prime}),
\end{equation}
where $\psi(\vec{x})$ is a two-component column vector, $\psi(\vec{x}%
)=(c^{\dagger}(\vec{x}), c(\vec{x}))^{\mathrm{T}}$, and ${\mathcal{H}}(\vec{x%
},\vec{x}^{\prime})$ is the matrix,
\begin{equation}
{\mathcal{H}}(\vec{x},\vec{x}^{\prime})=(\frac{-\nabla^2}{2m}-\mu)\delta(%
\vec{x}-\vec{x}^{\prime})\sigma_z + \frac{\Delta(\vec{x},\vec{x}^{\prime})}{2%
}\sigma^{+}-\frac{\Delta^{*}(\vec{x},\vec{x}^{\prime})}{2}\sigma^{-}
\label{Hamiltonian1}
\end{equation}
Here, $m$ is the fermion mass, $\mu$ is the chemical potential, and $\Delta(%
\vec{x},\vec{x}^{\prime})$ is the gap function. We take
$\hbar=k_B=1$ in
this paper. 
In a uniform $p_x+ip_y$ state, the gap function takes the form in
momentum
space, $\Delta(\vec{p})=\frac{\Delta_0}{p_F}(p_x +i p_y)$. 
For $\mu < 0$, the fall-off of the pair relative wave function
$g(\vec{x}_1-\vec{x}_2)$ is exponential (pairs are tightly bound),
whereas, for $\mu > 0$ , it is algebraic \cite{Read00}. Following
Ref.~\cite{Read00}, we identify the system to be in the
strong-pairing phase for $\mu < 0$ and in the weak-pairing phase
for $\mu > 0$, the two phases separated by a topological phase
transition \cite{Volovik}.
The weak-pairing phase supports zero-energy Majorana fermions at the
vortex cores. By explicitly constructing the bound state wave
function using the Bogoliubov - de Gennes (BdG) equations, below we
show that, for $\mu > 0$, the zero-mode quasiparticle is
self-hermitian (Majorana), and, as $\mu$ is tuned to negative
values, these modes disappear in the topologically trivial
strong-pairing phase. For $p_x+ip_y$ wave superconductors, analogous
BdG equations have been discussed in
Refs.~\onlinecite{Ashvin,Stone1,Stone2}.


To construct the eigenfunction of the Hamiltonian,
Eq.~\ref{Hamiltonian1}, for the zero-energy state, if any, in the
presence of a vortex, we model the vortex by assuming the gap
function to be zero inside a circular area of radius $\xi$
(coherence length). Outside this radius, the gap function takes
the form $\Delta(\vec{p})=\frac{\Delta_0}{p_F}\exp(i\theta/2)(p_x
-i p_y)\exp(i\theta/2)$, where the total order parameter phase,
$\theta$, rotates by $2\pi$ around the vortex with unit vorticity.
In polar coordinates $(\rho, \theta)$, for $\rho<\xi$, the BdG
equations take the form,
\begin{equation}
[-\frac{1}{2m}(\frac{\partial^2}{\partial \rho^2}+\frac{1}{\rho}\frac{%
\partial}{\partial\rho}+\frac{1}{\rho^2}\frac{\partial ^2}{\partial \theta^2}%
)-\mu]\sigma_z \phi(\vec{x})=0,  \label{insidecore}
\end{equation}
where $\phi(\vec{x})=(u(\vec{x}), v(\vec{x}))^{\mathrm{T}}$. For
the
zero-energy state, we take the angular momentum operator $l=-i\frac{\partial%
}{\partial \theta}$ to have eigenvalue zero. The remaining parts of Eq.~\ref%
{insidecore} imply just ordinary Bessel equations of order zero
\cite{Table} for both $u$ and $v$. Since one of the two
independent solutions is divergent at the origin, we find the
solution for $\phi$,
\begin{equation}
\phi(\rho)=AJ_0(\sqrt{2m \mu}\rho)\zeta,  \label{insidesolution}
\end{equation}
where $J_0$ is the Bessel function of the first kind of order
zero, $A$ is a constant, and $\zeta$ is a constant spinor.

For solutions with $\rho>\xi$, we note that the gap operator can
be written in polar coordinates as,
$-i\frac{\Delta_0}{p_F}\exp(-i\theta/2)(\frac{\partial}{\partial \rho}-\frac{%
i}{\rho}\frac{\partial}{\partial \theta})\exp(i\theta/2) =-i\frac{\Delta_0}{%
p_F}(\frac{\partial}{\partial \rho}+\frac{1}{2\rho})-i\frac{\Delta_0}{p_F}l$%
. Using this, and for zero angular momentum, the BdG equations for
the zero-energy state, on multiplication by $-2m\sigma_z$, can be
written as,
\begin{equation}
[(((\frac{\partial ^2}{\partial \rho^2}+\frac{1}{\rho}\frac{\partial}{%
\partial \rho})+2m\mu) - \frac{2m\Delta_0}{p_F}(\frac{\partial}{\partial \rho%
}+\frac{1}{2\rho})\sigma_y )]\phi(\rho)= 0.  \label{outsidecore1}
\end{equation}
The solutions to this equation which are well-behaved at
$\rho\rightarrow
\infty$ can be written as $\phi(\rho)=\chi(\rho)\exp(-\frac{\Delta_0}{v_F}%
\rho)(1, -i)^{\mathrm{T}}$, where $\chi(\rho)$ satisfies
\begin{equation}
[\frac{\partial^2}{\partial\rho^2}+\frac{1}{\rho}\frac{\partial}{\partial
\rho}+(2m\mu-\frac{\Delta_0^2}{v_F^2})]\chi(\rho)=0,
\label{outsidecore2}
\end{equation}
where $v_F=\frac{p_F}{m}$. This is again Bessel equation of order
zero. Since both solutions are well-behaved at infinity (they are
asymptotically sinusoidal), the general solution for $\phi(\rho)$
can therefore be written as,
\begin{equation}
\phi(\rho)=[BJ_0(\kappa \rho)+CY_0(\kappa \rho)]\exp(i\frac{\pi}{4}-\frac{\Delta_0}{v_F}%
\rho)(1, -i)^{\mathrm{T}},  \label{outsidesolution}
\end{equation}
where $Y_0$ is the Bessel function of the second kind of order zero,
$\kappa= \sqrt{2m\mu-\Delta_0^2/v_F^2}$, B, C are constants, and the
phase factor $e^{i\frac{\pi}{4}}$ is for equal distribution of phase
between $\phi$ and $\phi^{\dagger}$ (see below). Next, to get a
complete solution for the zero-energy state, one needs to match the
wave-function and its derivative at $\rho=\xi$, and also normalize
the function in all space. These conditions will provide three
equations for the three constants A, B, and C, which can then be
straightforwardly solved in terms of the known parameters. Once the
solution $\phi(\rho)$ and, in turn, $u(\vec{x}), v(\vec{x})$ are
known, the quasiparticle operator for the zero-energy state can be
written as
\begin{equation}
\gamma^{\dagger}_0 = \int d^2x
(u(\vec{x})c^{\dagger}(\vec{x})+v(\vec{x})c(\vec{x}))
\label{quasiparticle}
\end{equation}
For the overall phase choice $e^{i\frac{\pi}{4}}$ in
Eq.~\ref{outsidesolution}, one can see from
Eq.~\ref{quasiparticle} that $\gamma_0^{\dagger}=\gamma_0$ : the
zero energy state is a self-hermitian Majorana state.


To see what happens to the zero-energy state in the strong-pairing
phase, this phase can be accessed, in the spirit of
Ref.~\onlinecite{Read00}, by staying within the mean-field BCS
theory and taking $\mu < 0$. Ref.~\onlinecite{Read00} argued that
the edge of a vortex could be viewed as a wall separating vacuum
with $\mu$ large and negative inside the core from the condensate
outside. Therefore, in the weak-pairing condensate only, the edge
acts like a domain wall between strong (inside the core) and
weak-pairing phases, while, in the strong-pairing condensate,
nothing interesting should result in the core. By extending the
mean-field theory to $\mu < 0$ \cite{Ashvin, Sumanta1},
and replacing $\mu$ by $-|\mu|$,
Eq.~\ref{insidecore} (for zero angular momentum) and
Eq.~\ref{outsidecore2} now imply \textit{modified} Bessel
equations of order zero
for inside and outside the core, respectively. The solution
$\phi^{\prime}(\rho)$ is now given by only one of the two modified
Bessel functions of order zero in each case, $I_0(\sqrt{2m
|\mu|}\rho)$ for $\rho < \xi$, and $K_0(\kappa^{\prime}\rho)$
 for $\rho
> \xi$, since the other one is divergent in the relevant region \cite{Table}.
Here,
$\kappa^{\prime}=(2m|\mu|+\frac{\Delta_0^2}{v_F^2})^{\frac{1}{2}}$.
The corresponding constants multiplying the solutions drop out
when one matches the solutions and their derivatives at
$\rho=\xi$, and divide one equation by the other.
 For generic values of the parameters, the resulting equation does not have
a solution \cite{Sumanta1}, and therefore, we do not expect a
zero-energy state in the strong-pairing phase.

For the sake of completeness, we mention here, that even for a
spin-triplet superconductor with $S_z=0$, in the weak-pairing
phase, there are two zero-modes at the core of a vortex with
$N=1$, one for each quasiparticle spin. However, these modes are
not Majorana modes, since there is mixing of up and down spin
`c'-operators in the definition of the quasiparticle operator
analogous to Eq.~\ref{quasiparticle}. Moreover, in the presence of
any spin-flip scattering, these degenerate modes will mix and
split. This is why one has to consider the exotic $N=\frac{1}{2}$
vortices in these systems in an attempt to realize particles with
non-Abelian statistics \cite{Sumanta} .

\emph{Non-Abelian statistics and unitary operators in the Hilbert
space.} When the system is in the weak-pairing superfluid phase, a
dilute gas of vortices can be created. Suppose there are $2n$ such
vortices in the optical trap. Each vortex will have a zero-energy
Majorana fermion attached to the
core. For $2n$ vortices, there are $2n$ such fermions, which we denote by $%
\gamma_i$, where $i$ counts the vortices. The Majorana fermions
can be combined pairwise to create $n$ complex fermionic states,
$c_i=\gamma_{2i} + i\gamma_{2i-1}, c^{\dagger}_i=\gamma_{2i} -
i\gamma_{2i-1}$. Each one of these complex fermionic states can be
either occupied or unoccupied, giving rise to $2^{n}$-fold
degeneracy in the Hilbert space protected by the gap, $\omega_0
\sim \frac{\Delta_0^2}{\epsilon_F}$, where $\Delta_0$ is the
amplitude of the pairing gap and $\epsilon_F$ is the Fermi energy,
to the first excited state in the vortex core. Two states of a
representative qubit are identified with the absence ($|0\rangle$)
or the presence ($c^{\dagger}_i|0\rangle$) of a superfluid
quasiparticle in the fermionic state constructed from
$\gamma_{2i-1}$ and $\gamma_{2i}$. Note that the two states are
degenerate and are not directly associated with any particular
vortices. It is this non-locality that protects the qubits from
decoherence due to the environment, which acts through local
operators. For initialization of the qubits, note that creating
vortices in pairs from the vacuum will always put each pair in the
$|0\rangle$ state at zero temperature ($T$). At finite $T$, there
is always a non-zero probability that a fermionic quasiparicle
will end up on a vortex pair. Since we can read these
non-destructively (see below), we can correct or discard the
$|1\rangle$'s. We now briefly describe the physics \cite{Ivanov}
behind the unitary transformations in this space, induced by
braiding of the vortices around one another. These unitary
transformations can be fruitfully utilized for quantum
computation.

Using the property that $\gamma$'s carry odd charge with respect
to the
gauge field of a vortex with unit vorticity, that is, $\gamma\rightarrow-%
\gamma$ for a phase change of $2\pi$, it follows that, upon
interchange of two neighboring vortices $1$ and $2$, $\gamma_1
\rightarrow \gamma_2$, but $
\gamma_2 \rightarrow -\gamma_1$ \cite{Ivanov}. 
The unitary operator in the two-dimensional Hilbert space that
enforces this transformation is given by,
\begin{equation}
T_{1} = \exp(\frac{\pi}{4}\gamma_{2}\gamma_{1})=\exp(i\frac{\pi}{4}%
(2c^{\dagger}c-1)),  \label{operator}
\end{equation}
where, $c=\gamma_1+i\gamma_2$. This operator can be written as a
$2\times 2$ matrix in the space of states spanned by $|0\rangle$ and
$c^{\dagger}|0\rangle$. Likewise, for four vortices $\gamma_1,
\gamma_2, \gamma_3$ and $\gamma_4$, the unitary braiding operators
can be written as $4\times 4$ matrices in the space spanned by the
basis states $|0\rangle, c^{\dagger}_1|0\rangle,
c^{\dagger}_2|0\rangle$ and $c^{\dagger}_1c^{ \dagger}_2|0\rangle$,
where $c^{\dagger}_1=\gamma_1+i\gamma_2$, and
$c^{\dagger}_2=\gamma_3+i\gamma_4$. In the case of $2n$ vortices,
the braiding operators are $2^n\times 2^{n}$-dimensional matrices:
they form a matrix representation of the braid group in two
dimensions. Upon braiding of two vortices, an initial state, which
is now a $2^n$-dimensional vector in the space of degenerate states,
is multiplied by these matrices and gets transformed to another
vector in this space. It is these unitary transformations that can
be utilized to build unitary quantum gates, and this is the essence
of topological quantum computation. Note that $T$ needs to be kept
lower than $\omega_0$. For Feshbach resonance-superfluids, which are
near the BCS-BEC transition (recall that it is actually a transition
between the weak- and strong-pairing phases, unlike in the $s$-wave
case, in which it is merely a crossover), $\Delta_0 \sim
\epsilon_F$, and so $T$ should be kept much lower than $\epsilon_F$,
which is realizable in these systems \cite{Greiner}. For this method
to succeed, it is imperative that the vortices can be braided around
one another like independent particles, which requires a low density
of vortices, and their movement be quantum coherent, both of which
are achievable in optical traps. Below, we show that the atoms in
optical traps also offer a natural strategy for determining the
state of a qubit after a computation has been performed.

\emph{Reading out the states of the qubits.} A central question in
the above scheme is how to determine the state of a qubit after a
computation has been performed. The two states of the qubit, as we
described above, are distinguished by the presence or absence of a
superfluid quasiparticle at the complex fermionic state when two
vortices are fused together. However, since these are
quasiparticles of a superfluid, they are chargeless. Moreover,
since the Majorana fermions are spinless, the complex fermion,
which is a linear combination of two Majorana fermions, is also
spinless. Therefore, one does not expect an excess of charge or
spin due to the presence of a quasiparticle at the core of the
composite vortex. One may look for subtle differences in charge or
energy \emph{distribution} at the core due to the presence or
absence of a quasiparticle, but these may be experimentally
difficult to achieve.


A different approach, suitable for atomic (or molecular)
superfluids only, is to use the internal energy levels of the
atoms themselves. The basic point is that, if there is an unpaired
atom at the core of the composite vortex (the qubit is in the
state $c^{\dagger }|0\rangle $), photons from a laser can be
absorbed to excite the atom to an
appropriately chosen excited level. 
If there is no quasiparticle there (the qubit is in the state
$|0\rangle $), there will be no absorption at this frequency.
Note that, during this process, one might end up exciting Cooper
pairs from outside the core as well. However, to excite an atom
bound in a Cooper pair with another atom in an identical internal
state, one first needs to break the pair, costing an energy
$2\Delta _{0}$. Thus, from this process, photons can only be
absorbed at a frequency shifted by $2\Delta _{0}$. Since the
typical spontaneous emission rate, $\sim \mathcal{O}(2\pi \times
1$ MHz), in such a detection process is much larger than $\Delta
_{0}\sim 2\pi \times 11$ KHz \cite{Greiner}, this method can be
applied only through intermediate states which induce a much
larger energy splitting between paired and unpaired atoms. Here we
illustrate this reading out scheme using $^{40}K$ atoms, although
the technique is applicable to other species as well.

Suppose the atoms in the superfluid are in the $4^{2}S_{1/2}$
hyperfine ground state $\left\vert i\right\rangle \equiv
\left\vert
F=9/2,m_{F}=-7/2\right\rangle $ in the case of $p$-wave resonance \cite%
{Regal}. To determine whether there is an unpaired atom inside a
composite vortex, a two-photon Raman pulse is applied that
transfers the unpaired atom to another hyperfine state $\left\vert
j\right\rangle \equiv \left\vert F=7/2,m_{F}=-5/2\right\rangle $.
The frequency difference between the two Raman lasers is adjusted
to be resonant with the hyperfine splitting between states
$\left\vert i\right\rangle $ and $\left\vert j\right\rangle $ for
the unpaired atom, but has a $2\Delta _{0}$ detuning for paired
atoms due to the energy cost to break a pair. The lasers have
maximal intensities located at the core of the vortex and their
beam waist width $w\approx 1.5$ $\mu m$ is much smaller than the
typical distance ($\geq 10\mu m$) between vortices
\cite{Ketterle}, allowing individual access to the qubits. The
Rabi frequency of the Raman pulse is chosen to have a Gaussian
shape $\Omega =\Omega _{0}\exp \left( -\omega _{0}^{2}t^{2}\right)
$ ($-t_{f}\leq t\leq t_{f}$) to reduce the impact on paired atoms
\cite{Zhang}. For a set of parameters $\Delta _{0}=2\pi \times 11$
KHz, $\omega _{0}=\Delta _{0}/2$, and $\Omega _{0}=1.77\omega
_{0}$, $t_{f}=5/\omega _{0}$ \cite{Zhang}, we find that the
unpaired atom is completely transferred from state $\left\vert
i\right\rangle $ to $\left\vert j\right\rangle $ by the Raman
pulse, while the probability for the paired atoms to be excited to
state $\left\vert j\right\rangle $ is about $6\times 10^{-6}$ and
may therefore be neglected.

To obtain a cycling transition necessary for the detection of the
unpaired atom, $\pi $ Raman pulses are applied to transfer the
unpaired atom to state $\left\vert k\right\rangle \equiv
\left\vert F=9/2,m_{F}=9/2\right\rangle $. Because of large Zeeman
splitting between different magnetic sublevels, these Raman pulses
may be performed in a short
period (no longer than 100$\mu s$). A focused $\sigma ^{+}$%
-polarized detection laser resonant with the cycling transition
$\left\vert k\right\rangle \rightarrow \left\vert l\right\rangle
\equiv \left\vert 5^{2}P_{3/2}:F=11/2,m_{F}=11/2\right\rangle $ is
then applied to detect atoms at state $\left\vert k\right\rangle $.
Here we choose the $4S\rightarrow 5P$ instead of $4S\rightarrow 4P$
transition for the detection laser to obtain smaller diffraction
limit as well as smaller spontaneous decay rate. In the
experiment \cite{Regal}, the magnetic field $B\approx 200$ G for the $p$%
-wave Feshbach resonance, which yields an effective detuning
$\delta \approx 2\pi \times 170$ MHz for the paired atoms at state
$\left\vert i\right\rangle $. The ratio between the number of the
spontaneously emitted photons by paired and unpaired atoms is
estimated to be $\left( \Gamma /2\delta \right) ^{2}\approx
1.2\times 10^{-5}$, where $\Gamma \approx 2\pi \times 1.2$ MHz is
the decay rate for the excited state $\left\vert l\right\rangle $.
Therefore, in the detection process, the impact on paired atoms
may be neglected, and resonant fluorescence is observed if and
only if initially there is one unpaired atom inside the vortex at
state $\left\vert i\right\rangle $.
For this read-out scheme to succeed, the temperature needs to be
kept as low as possible, so that there are no other unpaired atoms
in the bulk around the vortex cores. However, such thermally
excited quasiparticles are expected to occur near the trap edges,
and so for lasers sufficiently focused on the vortices near the
center of the trap, one should be able to significantly suppress
the unwanted signals. Finally, we mention that the resonant
detection laser may be replaced with a resonant multiphoton
ionization process of the unpaired atom, yielding a single ionized
electron that can be detected with essentially unit efficiency
\cite{Raimond}. Similar analysis as above also yields negligible
impact on paired atoms. The advantage of the multiphoton
ionization is that the detection may be done in parallel for all
unpaired atoms in the sense that the electrons can be imaged on a
channel plate, where the detection or no detection of the
electrons is a parallel readout of all the qubits. However, such
process is destructive and the ionized atoms cannot be reused.

In conclusion, we have proposed to use the two-dimensional,
spin-polarized, $p_x+ip_y$ atomic resonance-superfluid in the
weak-pairing phase, potentially realizable in optical traps, in a
suitable hardware for topological quantum computation. We have
given a realistic read-out scheme for the topological qubits, a
major hurdle in this field, using the internal states of the
constituent atoms.

We thank M. Stone, A. Vishwanath, and V. Scarola for insightful
discussions. This work was carried out at the KITP, UCSB, and we
thank the organizers and the participants of the Workshop on
Topological Phases and Quantum Computation. This research is
supported in part by the National Science Foundation under Grant
No. PHY99-07949, and ARO-DTO, ARO-LPS, and NSF.

\vskip -6mm 

\end{document}